\documentclass[twocolumn,showpacs,preprintnumbers,
amsmath,amssymb,superscriptaddress]{revtex4-2}

\usepackage{graphicx}
\usepackage{dcolumn}
\usepackage{bm}
\usepackage{ulem}
\usepackage{varioref}
\usepackage{xr-hyper}
\usepackage{xcolor}
\usepackage{color}

\usepackage{hyperref}
\hypersetup{colorlinks = true,urlcolor = blue,linkcolor = blue,citecolor = blue
}

\usepackage[dvipsnames]{xcolor}
\definecolor{mygreen}{rgb}{0.0, 0.6, 0.0}
\definecolor{pjorange}{rgb}{0.8, 0.3, 0.0}
\definecolor{jlblue}{rgb}{0.2, 0.5, 0.7}

\graphicspath{{fig/}{./fig_07/}{.}}

\begin{document}

\title{Evidence for topological flat band magnetism in Co$_3$Sn$_2$S$_2$}

\author{Ola~Kenji~Forslund}
\email{ola.forslund@physics.uu.se}
\affiliation{Physik-Institut, Universität Zürich, Winterthurerstrasse 190, CH-8057 Zürich, Switzerland}
\affiliation{Department of Physics and Astronomy, Uppsala University, Box 516, SE-75120 Uppsala, Sweden}

\author{Andrzej~Ptok}
\email{aptok@mmj.pl}
\affiliation{Institute of Nuclear Physics, Polish Academy of Sciences, W. E. Radzikowskiego 152, PL-31342 Krak\'{o}w, Poland}

\author{Chun~Lin}
\affiliation{Physik-Institut, Universität Zürich, Winterthurerstrasse 190, CH-8057 Zürich, Switzerland}

\author{Aksel~Kobia\l{}ka}
\affiliation{Department of Physics and Astronomy, Uppsala University, Box 516, SE-75120 Uppsala, Sweden}

\author{Wan~Yu Chen}
\affiliation{Department of Applied Physics, KTH Royal Institute of Technology, SE-106 91 Stockholm, Sweden}

\author{Ugne~Miniotaite}
\affiliation{Department of Applied Physics, KTH Royal Institute of Technology, SE-106 91 Stockholm, Sweden}

\author{Martin~M{\aa}nsson}
\affiliation{Department of Applied Physics, KTH Royal Institute of Technology, SE-106 91 Stockholm, Sweden}

\author{Yasmine~Sassa}
\affiliation{Department of Applied Physics, KTH Royal Institute of Technology, SE-106 91 Stockholm, Sweden}

\author{Timur~Kim}
\affiliation{Diamond Light Source, Didcot OX11 0DE, United Kingdom}

\author{Cephise~Cacho}
\affiliation{Diamond Light Source, Didcot OX11 0DE, United Kingdom}





\author{Jian~Rui~Soh}
\affiliation{Quantum Innovation Centre (Q.InC), Agency for Science, Technology and Research (A*STAR), 2 Fusionopolis Way, Innovis \#08-03, Singapore 138634, Singapore}
\affiliation{Centre for Quantum Technologies, National University of Singapore, 3 Science Drive 2, Singapore 117543, Singapore}

\author{Chin~Shen~Ong}
\affiliation{Department of Physics and Astronomy, Uppsala University, Box 516, SE-75120 Uppsala, Sweden}

\author{Vladislav~Borisov}
\affiliation{Department of Physics and Astronomy, Uppsala University, Box 516, SE-75120 Uppsala, Sweden}

\author{Olle~Eriksson}
\affiliation{Department of Physics and Astronomy, Uppsala University, Box 516, SE-75120 Uppsala, Sweden}
\affiliation{Wallenberg Initiative Materials Science for Sustainability, Uppsala University,
75121 Uppsala, Sweden}

\author{Johan~Chang}
\affiliation{Physik-Institut, Universität Zürich, Winterthurerstrasse 190, CH-8057 Zürich, Switzerland}

\date{\today}

\begin{abstract}
\textbf{
Correlations and topology are central pillars in the study of emergent electronic phases. Topological flat bands near the Fermi level unite these concepts, providing a route towards correlated topological phenomena and unconventional magnetism. Despite theoretical predictions, momentum-resolved observation of a topological flat band at the Fermi level has remained elusive. Using angle resolved photoemission spectroscopy, we observe a flat band at the Fermi level within the topologically non-trivial electronic structure of the magnetic kagome Weyl semimetal Co$_3$Sn$_2$S$_2$. We further uncover a strong temperature dependence in which the flat band shifts across the Fermi level, accompanied by a reconstruction of the Fermi surface. Strikingly, this evolution occurs on the same temperature scale as previously reported anomalies in the magnetic response. Together, these results provide evidence for topological flat band magnetism in Co$_3$Sn$_2$S$_2$, establishing a microscopic setting in which topology, electronic correlations, and magnetism become intertwined.}
\end{abstract}


\maketitle

To realise new quantum phases with potential device applications, a key step is to bridge electronic topology with strong electron correlations. Correlated electron systems have revealed rich physics including magnetism, Mott insulators, and unconventional superconductivity~\cite{Basov2011}, while topological materials have introduced geometric concepts such as the Berry phase as foundational to electronic behavior~\cite{Xiao2010}. Yet, to date, electronic correlations and Berry curvature driven band topology have largely been studied in isolation, limiting our understanding of emergent phenomena at their intersection.

\begin{figure*}
\centering
\includegraphics[width=\textwidth]{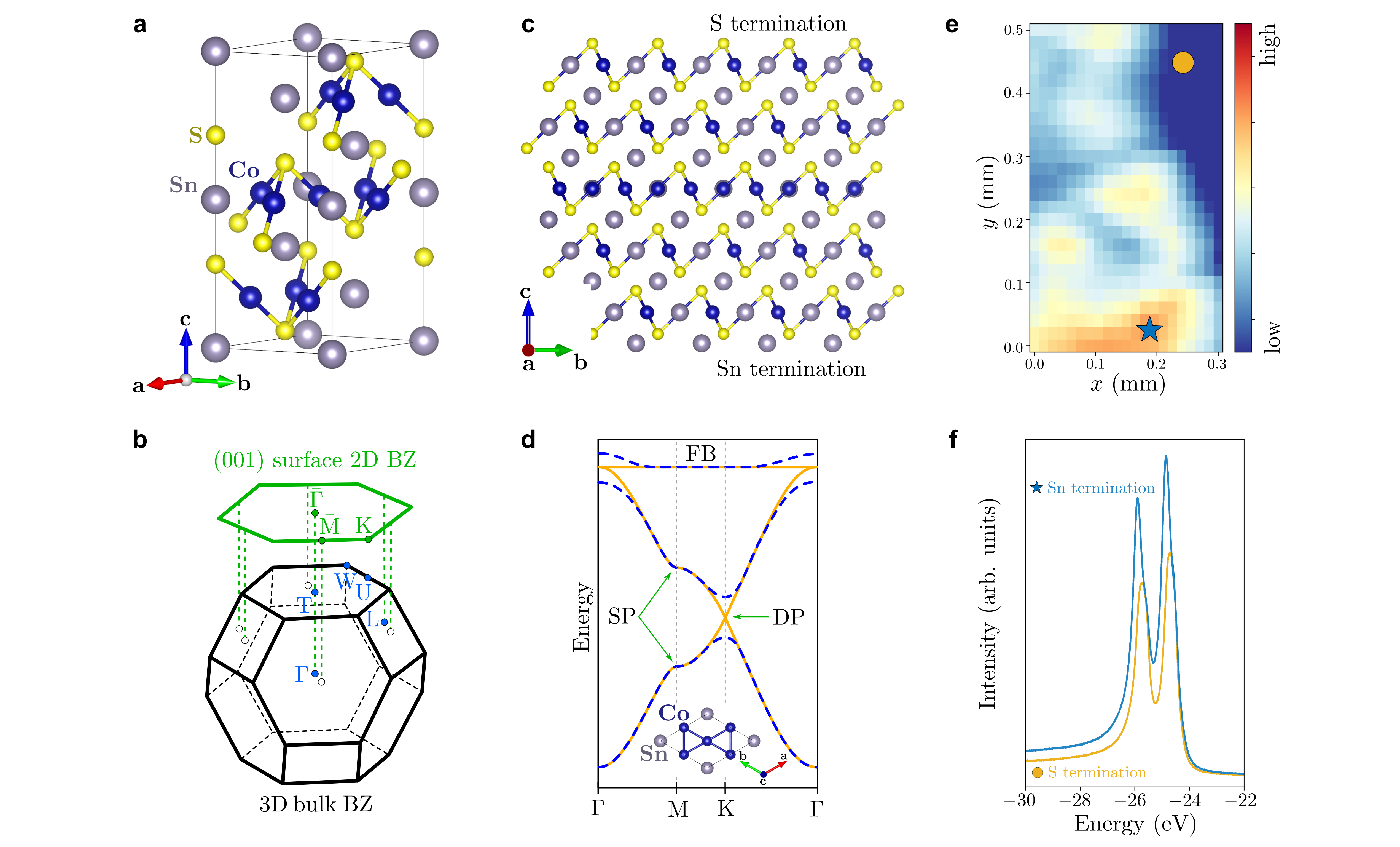}
\caption{
{\bf Structural properties of Co$_{3}$Sn$_{2}$S$_{2}$.}
(a) Conventional unit cell of Co$_3$Sn$_2$S$_2$. (b) The three-dimensional Brillouin zone and its projection onto the surface Brillouin zone. (c) Crystal structure with S or Sn termination for (001) surface (as labeled). (d) The electronic band structure of the ideal two-dimensional kagome lattice with (blue) and without (orange) spin-orbit coupling (SOC). Without SOC, the band structure contains a flat band (FB), saddle points (SP), and a Dirac point (DP). SOC gaps the Dirac crossing and gives rise to topologically non-trivial bands. The inset shows the kagome Co sublattice. (e) Raster scan X-ray photoelectron spectroscopy (XPS) measurement focused on the Sn core-level binding energy. The star and circle mark the positions corresponding to the spectra displayed in panel (f). (f) XPS spectra of Sn core-level excitations at S-terminated (yellow) and Sn-terminated (blue) surfaces.
\label{fig:figure1}}
\end{figure*}

A promising route to unify these frameworks lies in topological flat bands~\cite{Yin2022, Dumitru2025}. In such systems, electronic flatness coincides with non-trivial topology, strongly enhancing the density of states and interaction effects when the flat band approaches the Fermi level. This provides a route to correlated topological phases, including fractional anomalous quantum Hall states~\cite{Kai2011}, fractional topological insulators~\cite{Xiaohan2023}, topological superconductivity~\cite{Yuan2018}, and, in magnetic systems, topological flat band magnetism.
Magnetic kagome metals have emerged as leading candidates to realise this physics~\cite{Yin2022}. The kagome lattice naturally hosts flat bands through destructive interference in nearest-neighbor hopping models~\cite{Linda2024}. When combined with strong spin-orbit coupling (SOC), Dirac and band crossings near the K and $\Gamma$ points are gapped, rendering the flat bands topologically nontrivial (Fig.~\ref{fig:figure1}(d)). However, topological flat bands reported to date reside well below the Fermi level, excluding them from low energy physics~\cite{Zhiyong2018, Kang2020, meier.du.20, Zhenyu2022}.

The kagome system Co$_3$Sn$_2$S$_2$ is a magnetic Weyl semimetal that combines strong electronic correlations, topologically non-trivial bands, and unconventional magnetic behavior~\cite{Liu2019}. Angle resolved photoemission spectroscopy (ARPES) and theoretical work established its ferromagnetic Weyl semimetallic character, sparking extensive experimental and theoretical interest~\cite{Wang2018, Liu2019, Yin2019, jiao.xu.19, Yueshan2020, Belopolski2021, Liu2022, Mazzola2023}. Density functional theory (DFT) predicts flat band states, while scanning tunneling microscopy and optical studies have reported an enhanced momentum integrated density of states near the Fermi level consistent with flat band signatures~\cite{Yin2019, jiao.xu.19, Yueshan2020}. However, direct momentum-resolved observation of a flat band at the Fermi level has remained elusive~\cite{Wang2018, Liu2019, Belopolski2021, Liu2022, Mazzola2023}, leaving its possible connection to the unusual magnetic response unresolved. At the same time, differing ARPES observations concerning the role of SOC have complicated the interpretation of the low energy electronic structure.

Here, combining ARPES with DFT and dynamical mean field theory (DMFT), we resolve an extended flat band at the Fermi level within the topologically non-trivial electronic structure of Co$_3$Sn$_2$S$_2$. Calculations including dynamical electronic correlations reveal a pronounced renormalisation of the flat band states, highlighting the importance of correlations to the low energy electronic structure. Upon increasing temperature, the flat band shifts through the Fermi level, accompanied by a substantial reconstruction of the Fermi surface. Strikingly, this evolution occurs on the same temperature scale as previously reported magnetic anomalies. Together, these observations provide evidence for topological flat band magnetism in Co$_3$Sn$_2$S$_2$, establishing a microscopic setting in which topology, electronic correlations, and magnetism become intertwined.
\\~\\
%
\noindent\textbf{\label{sec:result}Surface terminations and spin--orbit coupling.} 
The Co$_3$Sn$_2$S$_2$ kagome lattice system is realised through Co atoms decorated with non-magnetic Sn atoms (Fig.~\ref{fig:figure1}(a)). This magnetic kagome layer is sandwiched between hexagonal S and Sn layers, giving rise to two surface terminations (Fig.~\ref{fig:figure1}(c)), as previously observed with STM~\cite{Yin2019, jiao.xu.19}. We confirm this finding by presenting raster scan X-ray photoelectron spectroscopy (XPS) measurements at the Sn core-level binding energy (Fig.~\ref{fig:figure1}(e, f)). 

\begin{figure*}
\centering
\includegraphics[width=\textwidth]{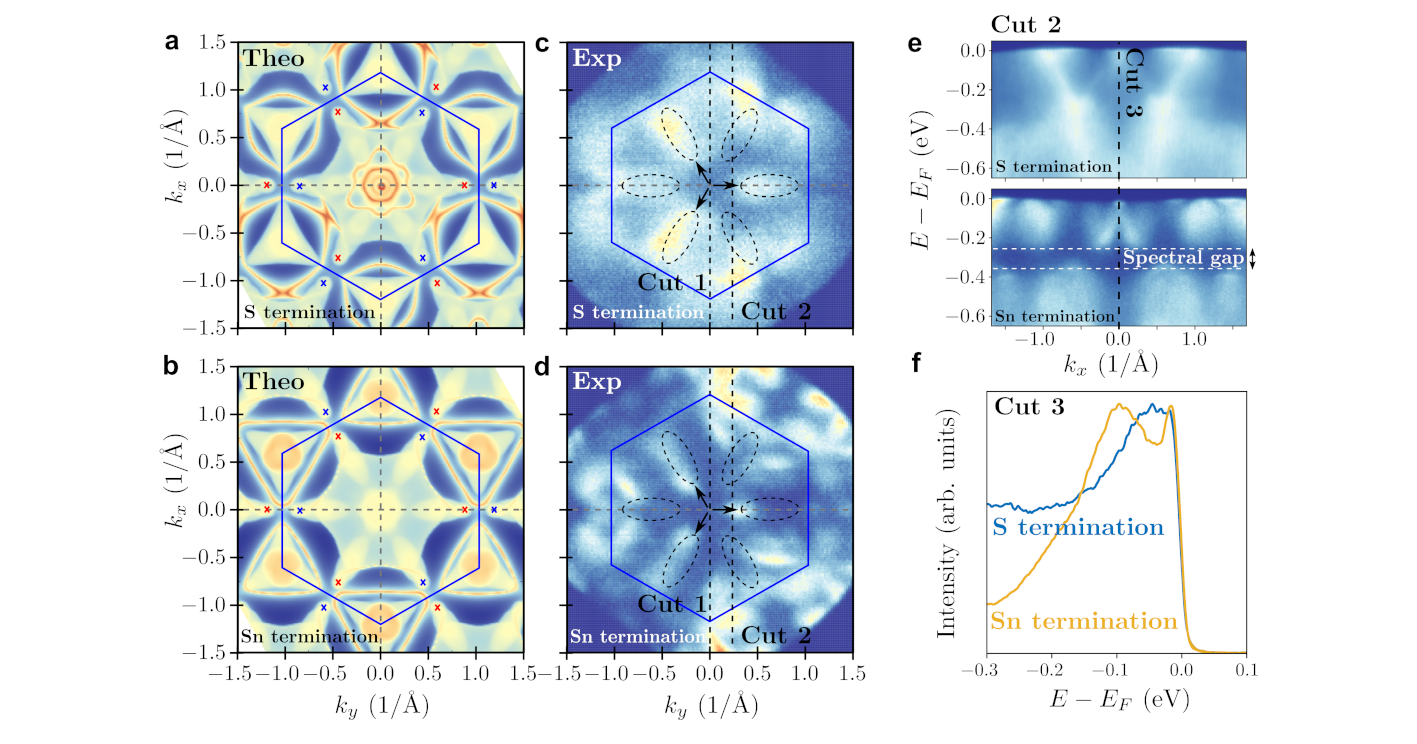}
\caption{
{\bf Terminations and spin-orbit gap.} 
(a-d) DFT calculated and experimental Fermi surfaces for S and Sn termination (top and bottom panels, respectively). (e) Momentum distribution cut parallel to $\bar{\Gamma}$-$\bar{\text{K}}$ (cut~2 as indicated in (c, d)) for S (top) and Sn (bottom) terminations. (f) An energy distribution curve for both terminations as indicated in (e; cut~3). The colour scale is defined in Fig.~\ref{fig:figure1}(e).
\label{fig:figure2}}
\end{figure*}

Figure~\ref{fig:figure2}(c, d) displays the measured Fermi surface for each termination, using a photon energy $h \nu = 130$~eV at $T = 21$~K. Photon energy scans presented in supplementary materials (SM; Fig.~S1) verify that $130$~eV corresponds approximately to the bulk L~point~\cite{Liu2019,Belopolski2021,Liu2022}. In this geometry, bulk state contributions project onto the 2D Brillouin zone such that the T--W and T--U bulk directions map onto the $\bar{\Gamma}$-$\bar{\text{K}}$ and $\bar{\Gamma}$-$\bar{\text{M}}$ directions, respectively (Fig.~\ref{fig:figure1}(b)). The measured Fermi surfaces (Fig.~\ref{fig:figure2}(c,d)) confirm strong variation in spectral features depending on the surface termination~\cite{Mazzola2023}. This is consistent with orbital contributions inferred from prior~\cite{Huang2023} and our own Fermi surface calculations from DFT (Fig.~\ref{fig:figure2}(a,b)).



Consistent with prior ARPES work~\cite{Wang2018, Liu2019, Belopolski2021, Liu2022, Mazzola2023}, three point-like features are observed at the Fermi level around $\bar{\Gamma}$, marking the edges of elliptical nodal line pockets (black arrows and dashed lines in Fig.~\ref{fig:figure2}(c,d)). Yet, the role of SOC in Co$_3$Sn$_2$S$_2$ has been debated~\cite{Belopolski2021, Liu2022}. Cuts through these
features (Fig.~\ref{fig:figure2}(e,f)) resolve this by revealing a clear SOC-induced gap for Sn-terminated \cite{Liu2022} surfaces but not for S-termination. This difference, based on different terminations, explains the prior conflicting reports~\cite{Belopolski2021, Liu2022}.

\begin{figure*}
\centering
\includegraphics[width=\textwidth]{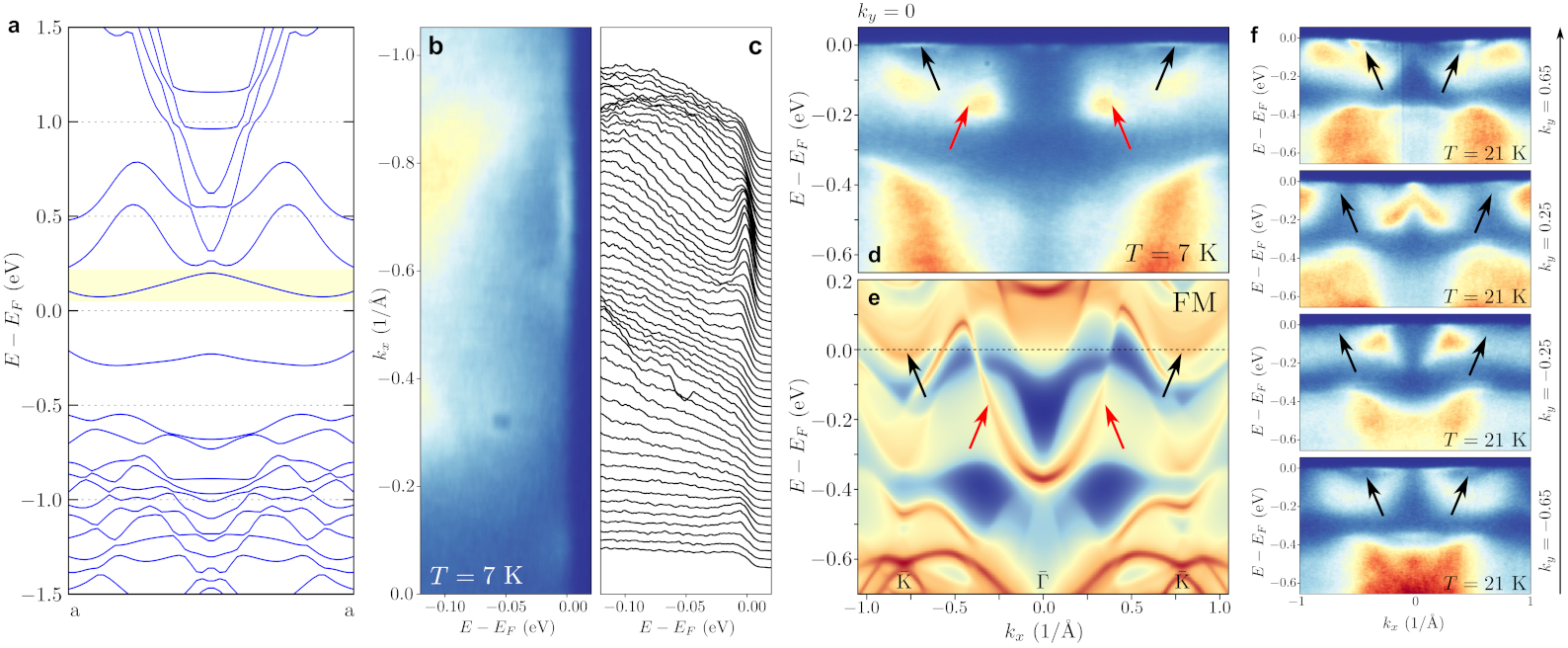}
\caption{
{\bf Flat bands around the Fermi level.} 
(a) Calculated electronic band structure along the path shown in Fig.~S6. The yellow area highlights the experimentally observed flat band. (b, c) Close up view of the ARPES spectrum in panel (d) near the Fermi level, together with the corresponding EDC cuts. (d, e) ARPES spectra along $\bar{\Gamma}$--$\bar{\text{K}}$ direction (cut 1 in Fig.~\ref{fig:figure2}(d)) and the corresponding theoretically calculated spectral function for Sn termination. Experimentally observed flat bands at the Fermi level are highlighted with black arrows. The deeper experimental bands around $\bar{\Gamma}$ can be recognised as parts of the theoretically obtained parabolic-like band, shown by red arrows. (f) Series of cuts parallel to $\bar{\Gamma}$--$\bar{\text{K}}$ for the specified $k_y$. The arrows highlight flat bands observed in experiment at the Fermi level. The colour scale is defined in Fig.~\ref{fig:figure1}(e).
\label{fig:figure3}
}
\end{figure*}

\begin{figure*}
\centering
\includegraphics[width=\textwidth]{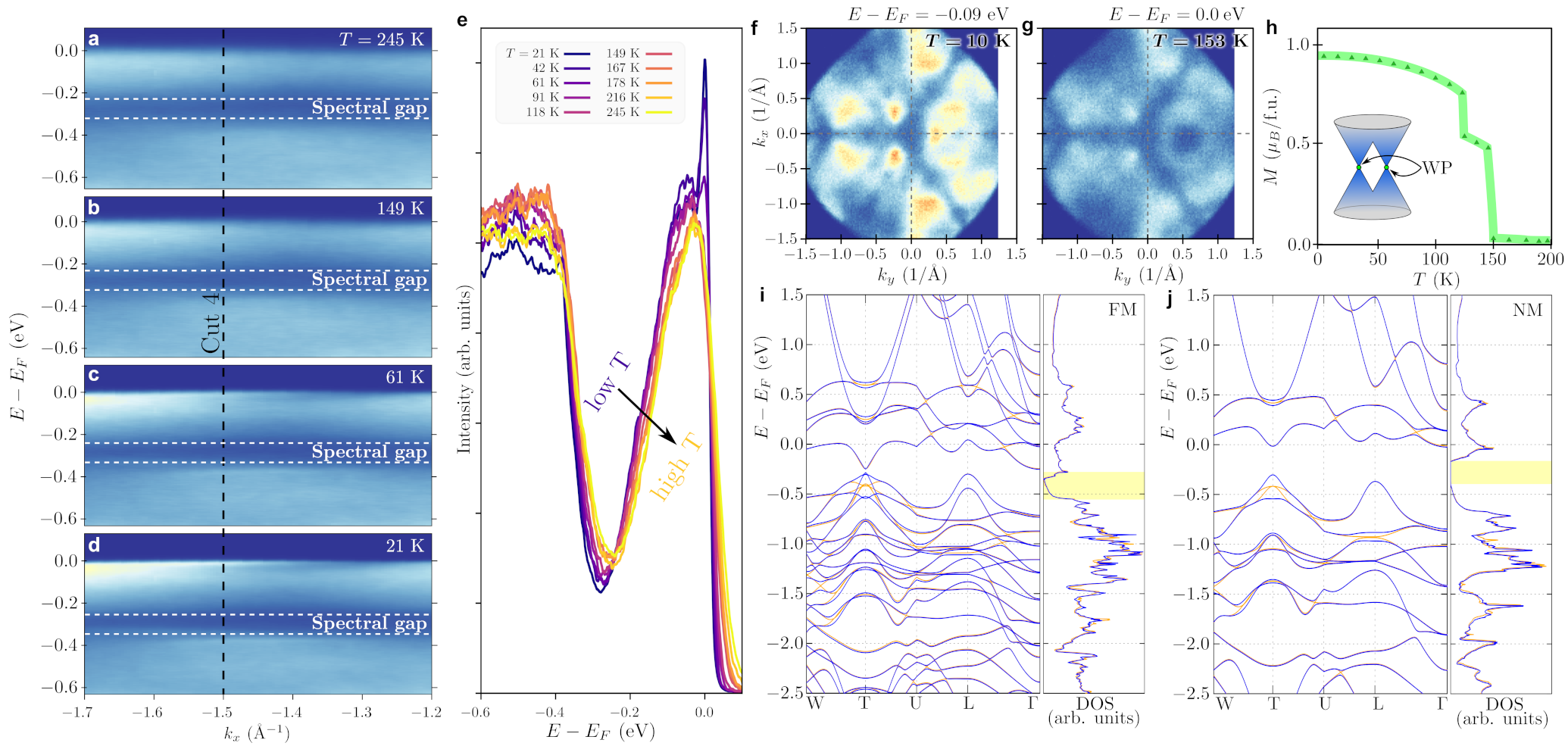}
\caption{
{\bf Temperature dependence of ARPES spectra.}
(a--d) Cuts parallel to $\bar{\Gamma}$--$\bar{K}$ (cut 2 in Fig.~\ref{fig:figure2}(d)) collected at different temperatures (as labeled). The dashed horizontal lines indicate the spectral gap. (e) Corresponding energy distribution cuts across the dashed lines in (a--d; cut 4) panels as a function of temperature as indicated. (f) The measured constant energy surface below the Fermi level ($E - E_{\rm F} = -0.09$~eV) collected at $T= 10$~K. (g) The measured Fermi surface collected at $T = 153$~K (cf. Fig.~\ref{fig:figure2}(d)). (h) Magnetisation as a function of temperature, obtained after field-cooling along the $c$-axis and measured on warming in zero field. Data reproduced from Ref.~\cite{Soh2022}. The inset highlights the existence of Weyl points (WPs) in the magnetic phase. (i,j) The electronic band structure and density of states (DOS) in the ferromagnetic and non-magnetic states, with (blue) and without (yellow) spin-orbit coupling. The colour scale is defined in Fig.~\ref{fig:figure1}(e).
\label{fig:figure4}}
\end{figure*}

Another termination effect is the emergence of a large spectral gap ($\sim150$~meV) centered around $-0.3$~eV on the Sn-terminated surface (Fig.~\ref{fig:figure2}(e), Fig.~\ref{fig:figure3}(d-f), Fig.~\ref{fig:figure4}(a-d)). Here, we employ the term "spectral gap" to describe a region in which no band crossings are observed experimentally, whereas the calculated density of states (DOS) is small but non-zero. This spectral gap persists throughout the 3D momentum space (SM; Fig.~S2), and is reflected in the calculated DOS (Fig.~\ref{fig:figure4}(i)). In contrast, this spectral gap is absent for the S-terminated surface.
\\~\\
{\bf Flat band.}
Theoretical DFT calculations predict a flat band along the high-symmetry $\text{a}$-$\text{a}$ direction (Fig.~\ref{fig:figure3}(a); SM Fig.~S6). 
In agreement with this prediction, ARPES measurements taken at $h\nu=130$~eV on the Sn terminated surface along this direction (Fig.~\ref{fig:figure3}(d)) reveal a flat band near the Fermi level (Fig.~\ref{fig:figure3}(a)). Similar features are also observed on the S-terminated surface (Fig.~S3(a)), suggesting that the flat band is not specific to a particular surface termination and is consistent with a bulk-derived origin. The presence of these bands is further emphasised in the second derivative plots shown in Fig.~S3. This observation aligns with prior optical and STM studies that suggested the presence of flat bands in this system~\cite{Yin2019, jiao.xu.19, Yueshan2020}.

The characteristics of these experimentally observed bands are highlighted through a series of cuts near the Fermi level as a function of $k_x$ (Fig.~\ref{fig:figure3}(b,c)). The resulting enhancement of the DOS provides a clear spectroscopic signature of flat band formation. The flat band exhibits a bandwidth of approximately 18~meV across the Brillouin zone, indicating that electron kinetic energy is strongly quenched due to destructive interference that suppresses wavefunction delocalisation. Experimentally, the band remains flat over an extended momentum range of at least 0.6~\AA$^{-1}$ (Fig.~\ref{fig:figure3}(b,c)), corresponding to roughly 75\% of the Brillouin zone in this direction, even as its spectral weight diminishes near the Fermi level. We further confirm that the flat band extends along $k_y$ (Fig.~\ref{fig:figure3}(f)), while similar flat band spectral weight persists over a photon energy range, corresponding to different $k_z$ values (SM; Fig.~S2). However, the experimentally observed intensity is not uniformly distributed: the flat band spectral weight appears in separated momentum patches that collectively cover about 50\% of the Brillouin zone. 

To better understand this behavior in Co$_3$Sn$_2$S$_2$, we compute the spectral function along $\bar{\Gamma}$--$\bar{\text{K}}$ (Fig.~\ref{fig:figure3}(e)), incorporating both bulk and surface-state contributions. While the calculations qualitatively capture the overall band structure, particularly at higher binding energies, they significantly underestimate the extent of the flat band near the Fermi level. This discrepancy is consistent with a recent study showing that calculations for Co$_3$Sn$_2$S$_2$ generally fail to fully reproduce experimental spectral features~\cite{Sudheer2025}. Such differences were attributed to the strong sensitivity of the surface electronic structure to local termination and disorder, which are not captured by idealized slab calculations. Nevertheless, the computed band (Fig.~\ref{fig:figure3}(a,e)) exhibits a small upward dispersion -- similar to the observed flat band in other kagome systems~\cite{Kang2020, Zhenyu2022} -- suggesting that portions of the experimentally observed flat band lie above the Fermi level. This dispersion combined with matrix element effects likely leads to the observed intensity distribution. 



The emergence of flat bands in Co$_3$Sn$_2$S$_2$ is particularly intriguing in light of its established~\cite{Liu2019,Guguchia2020,Mazzola2023} topological character. Theoretical analysis~\cite{sun.liu.23} of the Chern number reveals that $k_z$-planes away from the Weyl nodes exhibit enhanced topological character with large Chern numbers ($C=-6$ and $C=-9$)~\cite{sun.liu.23}. The flat band observed in our data is located within these topologically non-trivial regions of momentum space, placing it within the strongly topological low energy electronic structure of Co$_3$Sn$_2$S$_2$. This conclusion aligns with transport measurements showing pronounced anomalous Hall conductivity arising from Berry curvature near Fermi level~\cite{Liu2018, Wang2018}.
%
%
\\~\\
{\bf Temperature dependence.}
With the flat bands established at the Fermi level, we next examine their temperature evolution. In Co$_3$Sn$_2$S$_2$, magnetic order strongly influences the low-energy electronic structure through the exchange splitting and spin polarisation of states predominantly derived from Co-$d$ orbitals. As the temperature increases toward the Curie temperature ($T_C$), the bulk magnetisation progressively decreases (Fig.~4(h)), reducing the exchange splitting and modifying the spin polarisation of states near the Fermi level. This evolution of the magnetic order is therefore expected to substantially reorganise the low-energy electronic structure.

These effects are reflected in the electronic structure of Co$_3$Sn$_2$S$_2$ (Fig.~\ref{fig:figure4}(a-e)), as revealed by temperature dependent cuts along $\bar{\Gamma}$-$\bar{\text{K}}$ (Fig.~\ref{fig:figure4}(a-e)) on the Sn terminated surface. The flat bands progressively shift in energy and cross the Fermi level above $100$~K (Fig.~\ref{fig:figure4}(a-d)), reflecting a total shift of about $100$~meV. This shift is further illustrated by comparing the high temperature Fermi surface map with the low temperature constant energy surface at $E - E_F = -90$~meV (Fig.~\ref{fig:figure4}(f,g)). Even though the intensity is different, the surfaces are remarkably similar. Evidence of these shifts is also apparent in the EDC curves (Fig.~\ref{fig:figure4}(e), Fig.~S4), where peaks near the Fermi level move above it while the incoherent background decreases, consistent with a rigid band shift rather than a loss of coherent peak intensity. These results are consistent with the contrast between the calculated ferromagnetic and non-magnetic electronic structures (Fig.~\ref{fig:figure4}(i,j)).

This temperature induced band shift has two notable consequences. First, the electronic structure exhibits a clear Fermi-surface reconstruction, as evident from the high- and low-temperature Fermi surfaces (Fig.~\ref{fig:figure4}(g) and Fig.~\ref{fig:figure2}(d)). Second, the crossings of flat bands and nodal lines around $T \simeq 100$~K correlate closely with previously reported magnetic anomalies~\cite{Guguchia2020, Soh2022, ekahana.okamoto.24}. Unlike nonmagnetic flat band systems, where temperature primarily induces thermal broadening or minor energy shifts~\cite{Sutter2019}, the flat band observed here shifts in energy. These results link the temperature dependent band reconstruction in Co$_3$Sn$_2$S$_2$ to magnetic order, providing evidence for topological flat band magnetism at low temperatures.
\\~\\
\noindent\textbf{Electronic correlations.}
To further examine the role of electronic correlations, we performed bulk DMFT calculations including dynamical correlations of the Co-$d$ states. Compared with the corresponding DFT calculation, electronic correlations lead to a narrowing of several low energy bands and a pronounced renormalisation of flat band features (Fig.~S7). Spin- and orbital-resolved calculations further show that the calculated bands at the Fermi level are strongly spin-polarised and predominantly derived from Co-$d$ states (Fig.~S8). This close connection to the magnetic Co-$d$ electronic structure, together with the correlation-induced band renormalisation, indicates that the low energy electronic structure is sensitive to both magnetic order and dynamical electronic correlations, consistent with the pronounced temperature-dependent band evolution observed experimentally in Fig.~\ref{fig:figure4}.
\\~\\
\noindent\textbf{\label{sec:discussion}Topological flat band magnetism.} Magnetic probes, including neutron diffraction, bulk magnetisation, and muon spin relaxation, have reported anomalous magnetic responses in Co$_3$Sn$_2$S$_2$ around $T\sim90 -120$~K~\cite{Guguchia2020, Soh2022, ekahana.okamoto.24}, coinciding with a significant modification in the anomalous Hall conductivity~\cite{Soh2022}. However, the microscopic origin of these anomalies has remained unclear. Recent neutron and Mössbauer measurements identify a canted, noncollinear ferromagnetic state that persists throughout the magnetically ordered regime~\cite{Tam2026}, further arguing against these anomalies originating from a transition between distinct bulk magnetic phases. Instead, our ARPES measurements provide a direct momentum--energy resolved perspective, revealing that these magnetic anomalies coincide with a reconstruction of the Fermi surface. The overall band shift evolves on the same temperature scale as the temperature dependent magnetisation, linking the observed reconstruction to magnetic order. Notably, this reconstruction involves the observed flat band crossing the Fermi level.


Taken together with the correlation-induced renormalisation and strong spin polarisation of the flat band states revealed by our calculations, these observations provide evidence for topological flat band magnetism in Co$_3$Sn$_2$S$_2$. They establish a microscopic setting in which topology, electronic correlations, and magnetism become intertwined through the energetic position of a flat band. Establishing whether the flat band itself drives the anomalous magnetic response will require experiments that controllably tune its energetic position while directly tracking the corresponding magnetic response.%
%
%
%
\\~\\
%
%
\\~\\
{\bf Methods}
\\
{\bf Crystal synthesis.} Single crystals of Co$_3$Sn$_2$S$_2$ were grown at the Magnetism and Magnetic Materials Laboratory at Kyoto University using a Sn flux method following Ref.~\cite{KASSEM2015208}. High-purity Co (99.9\% High Purity Chemicals), Sn (99.999\% High Purity Chemicals), and S powder (99.999\% Nacalai Tesque) were mixed in a molar ratio of Co:S:Sn = 8:6:86, with a total batch mass of $\sim$15\,g. The mixture was loaded into an Al$_2$O$_3$ growth crucible, which was capped with an inverted Al$_2$O$_3$ crucible filled with quartz wool. Both were sealed under vacuum in a quartz tube. The ampule was heated to 1050$^\circ$C for 6 hrs, then slowly cooled to 700$^\circ$C over 72 hrs, before removal from the furnace. Excess Sn flux was removed by centrifuging the decanted ampule, yielding plate-like crystals. Powder x-ray diffraction (XRD) and wavelength dispersive x-ray spectroscopy (WDX) confirmed the single-phase nature and stoichiometric composition of Co$_3$Sn$_2$S$_2$. Laue x-ray diffraction further verified the high crystalline quality and confirmed that the broad crystal faces were parallel to the $ab$ plane.
\\~\\
{\bf Angle resolved photoemission spectroscopy (ARPES).} 
The ARPES measurements were performed at the I05 beamline at Diamond Light Source as a function of temperature. The beam spot size was about $50 ~\mu$m. The crystals were cleaved in-situ using the toppost method, and measured in ultra high vacuum of $\sim 2 \times 10^{-10}$~mbar. The pressure was kept below $\sim 5 \times 10^{-10}$~mbar during the temperature cycling. All Fermi surfaces were mapped using deflector mode of the electron analyser. ARpest was used to analyse the ARPES data~\cite{ARpest2025}. 
\\~\\
{\bf Computational details.}
The first-principles density functional theory (DFT) calculations were carried out using the projector augmented-wave (PAW) potentials~\cite{blochl.94} as implemented in the \textit{Vienna Ab initio Simulation Package} ({\sc Vasp})~\cite{kresse.hafner.94,kresse.joubert.99}. The exchange-correlation energy was treated using the generalised gradient approximation (GGA) within the Perdew--Burke--Ernzerhof (PBE) parametrisation~\cite{perdew.burke.96}. 
An energy cutoff of $350$~eV was applied for the plane-wave expansion. The van der Waals corrections were introduced within DFT-D3 method with Becke-Johnson damping function~\cite{grimme.ehrlich.11}. Calculations were performed in absence and presence of the spin--orbit coupling (SOC). Lattice constants and atomic positions were optimised for conventional unit cells with ferromagnetic order, containing three formula units. The optimisation was performed using a $13 \times 13 \times 5$ \textbf{k}-point grid within the Monkhorst--Pack scheme~\cite{monkhorst.pack.76}. The convergence criteria for the optimisation loop were set to an energy change below $10^{-6}$~eV for the ionic degrees of freedom and $10^{-8}$~eV for the electronic degrees of freedom. 

The symmetry of the structures was analysed using {\sc FindSym}~\cite{stokes.hatch.05} and {\sc Spglib}~\cite{togo.tanaka.18}, while momentum space analysis was conducted with {\sc SeeK-path}~\cite{hinuma.pizzi.17}. After system optimisation, we obtain lattice constants $a = 5.300$~\AA, and $c = 12.984$~\AA, which is close to the experimental values~\cite{Huang2023,jiao.xu.19}.
Atoms are located at the Wyckoff positions Co $9d$ (1/2,0,1/2), Sn(1) $3a$ (0, 0, 0), Sn(2) $3b$ (0, 0, 1/2), and S $6c$ (0, 0, 0.7174). The calculated Co magnetic moment is $0.34~\mu_\text{B}$, consistent with experimental results~\cite{Wang2018,xing.shen.20}.

The tight-binding model in the basis of maximally localised Wannier orbitals~\cite{marzari.mostofi.12,souza.marzari.01} was derived from the exact DFT electronic band structure using the {\sc Wannier90} software~\cite{mostofi.yates.08,pizzi.vitale.20}. In our calculations, a $\Gamma$-centered \textbf{k}-point grid of $12 \times 12 \times 12$ was employed, with the initial projections chosen as the $s$ and $d$ for Co, while $s$ and $p$ orbitals for Sn and S atoms. Furthermore, this model was applied to study the surface Green's function for a semi-infinite system using the {\sc WannierTools} package~\cite{sancho.sancho.85,wu.zhang.18}.

In addition to the VASP-based calculations described above, the bulk electronic structure of Co$_3$Sn$_2$S$_2$ was calculated using the full-potential Linear Muffin-Tin Orbital RSPt code~\cite{Wills1987,Wills2010}, within the GGA-PBE approximation~\cite{perdew.burke.96}. To investigate the effect of dynamical electronic correlations, calculations were performed within dynamical mean-field theory (DMFT). The dynamical correlations of the Co-$d$ states were modelled within the FLEX approximation, using $U=3$~eV and $J=0.9$~eV. This approximation is well established for moderately correlated metallic systems where the correlation energy is smaller than approximately half the bandwidth.
\\~\\
{\bf Data availability}
\\
The data that support the findings in this study are available from the corresponding author upon reasonable request.
\bibliography{Refs} 
%
%
{\bf Acknowledgments}
\\
This work was carried out with the support of Diamond Light Source, instrument I05. O.K.F. is supported by the Swedish Research Council (Grant No. 2022-06217), an \AA{}Forsk repatriation grant (Ref. No. 25-1), and the Ruth and Nils-Erik Stenb\"ack Foundation. U.M., M.M., and Y.S. acknowledge funding from the Swedish Research Council, VR (Dnr. 2021-06157 and Dnr. 2022-03936), the Swedish Foundation for Strategic Research (SSF) within the Swedish national graduate school in neutron scattering (SwedNess), the Carl Tryggers Foundation for Scientific Research (CTS-22:2374), and the Knut and Alice Wallenberg Foundation through the grant 2021.0150. V.B. acknowledges support by the Swedish Research Council through grant no.~2024.05206. Some figures in this work were rendered using {\sc Vesta}~\cite{momma.izumi.11} software. 
\\~\\
{\bf Author contributions}
\\
The project was conceived by O.K.F. The ARPES experiments were carried out by O.K.F, C.L, W.Y.C, U.M, M.M, Y.S, T.K, C.C, J.R.S and J.C. A.P, A.K, C.S.O, V.B and O.E conducted the theoretical calculations. O.K.F and A.P prepared the manuscript, and all co-authors contributed to the final draft.
\\~\\
{\bf Competing interests}
\\
The authors declare no competing interests.
\end{document}